\documentclass[aps,reprint,amssymb,prd,floatfix,superscriptaddress,nofootinbib]{revtex4-1}

\usepackage[english]{babel}
\usepackage[utf8]{inputenc}
\usepackage{comment}
\usepackage[colorinlistoftodos]{todonotes}
\usepackage{amsmath}
\usepackage{amsthm}
\usepackage{amsbsy}
\usepackage{amssymb}
\usepackage{latexsym}
\usepackage{graphicx}
\usepackage{epstopdf}
\usepackage[caption=false]{subfig}

\makeatletter
\let\@fnsymbol\@alph
\makeatother

\begin{document}
\captionsetup[subfigure]{labelformat=empty}

\title{Search for dark matter decay of the free neutron from the UCNA experiment: \\ n $\rightarrow \chi + e^+e^-$}

\author{X.~Sun}                          \affiliation{W.~K.~Kellogg Radiation Laboratory, California Institute of Technology, Pasadena, California 91125, USA}

\author{E.~Adamek}                       \affiliation{Department of Physics, Indiana University, Bloomington, Indiana 47408, USA}
\author{B.~Allgeier}                     \affiliation{Department of Physics and Astronomy, University of Kentucky, Lexington, Kentucky 40506, USA}
\author{M.~Blatnik}                      \affiliation{W.~K.~Kellogg Radiation Laboratory, California Institute of Technology, Pasadena, California 91125, USA}

\author{T.~J.~Bowles}                    \affiliation{Los Alamos National Laboratory, Los Alamos, New Mexico 87545, USA}
\author{L.~J.~Broussard}           \thanks{Currently at Oak Ridge National Laboratory, Oak Ridge, TN 37831, USA} \affiliation{Los Alamos National Laboratory, Los Alamos, New Mexico 87545, USA}
\author{M.~A.-P.~Brown}  \thanks{Currently at Humana Military, 305 N. Hurstbourne Pkwy, Louisville, KY 40222, USA}                \affiliation{Department of Physics and Astronomy, University of Kentucky, Lexington, Kentucky 40506, USA}
\author{R.~Carr}                         \affiliation{W.~K.~Kellogg Radiation Laboratory, California Institute of Technology, Pasadena, California 91125, USA}
\author{S.~Clayton}                      \affiliation{Los Alamos National Laboratory, Los Alamos, New Mexico 87545, USA}
\author{C.~Cude-Woods}                   \affiliation{Department of Physics, North Carolina State University, Raleigh, North Carolina 27695, USA}
\author{S.~Currie}                       \affiliation{Los Alamos National Laboratory, Los Alamos, New Mexico 87545, USA}
\author{E.~B.~Dees}                      \affiliation{Department of Physics, North Carolina State University, Raleigh, North Carolina 27695, USA} \affiliation{Triangle Universities Nuclear Laboratory, Durham, North Carolina 27708, USA}
\author{X.~Ding}                         \affiliation{Department of Physics, Virginia Tech, Blacksburg, Virginia 24061, USA}
\author{B.~W.~Filippone}                 \affiliation{W.~K.~Kellogg Radiation Laboratory, California Institute of Technology, Pasadena, California 91125, USA}
\author{A.~Garc\'{i}a}                   \affiliation{Department of Physics and Center for Experimental Nuclear Physics and Astrophysics, University of Washington, Seattle, Washington 98195, USA}
\author{P.~Geltenbort}                   \affiliation{Institut Laue-Langevin, 38042 Grenoble Cedex 9, France}
\author{S.~Hasan}                        \affiliation{Department of Physics and Astronomy, University of Kentucky, Lexington, Kentucky 40506, USA}
\author{K.~P.~Hickerson}                 \affiliation{W.~K.~Kellogg Radiation Laboratory, California Institute of Technology, Pasadena, California 91125, USA}
\author{J.~Hoagland}                     \affiliation{Department of Physics, North Carolina State University, Raleigh, North Carolina 27695, USA}
\author{R.~Hong}                         \affiliation{Department of Physics and Center for Experimental Nuclear Physics and Astrophysics, University of Washington, Seattle, Washington 98195, USA}
\author{G.~E.~Hogan}                     \affiliation{Los Alamos National Laboratory, Los Alamos, New Mexico 87545, USA}
\author{A.~T.~Holley}                    \thanks{Currently at Dept. of Physics, Tennessee Tech University, Cookeville, TN, USA}\affiliation{Department of Physics, North Carolina State University, Raleigh, North Carolina 27695, USA} \affiliation{Department of Physics, Indiana University, Bloomington, Indiana 47408, USA}
\author{T.~M.~Ito}                       \affiliation{Los Alamos National Laboratory, Los Alamos, New Mexico 87545, USA}
\author{A.~Knecht}                       \thanks{Currently at Paul Scherrer Institut, 5232 Villigen PSI, Switzerland} \affiliation{Department of Physics and Center for Experimental Nuclear Physics and Astrophysics, University of Washington, Seattle, Washington 98195, USA}

\author{C.-Y.~Liu}                       \affiliation{Department of Physics, Indiana University, Bloomington, Indiana 47408, USA}
\author{J.~Liu}                          \affiliation{Department of Physics, Shanghai Jiao Tong University, Shanghai, 200240, China}
\author{M.~Makela}                       \affiliation{Los Alamos National Laboratory, Los Alamos, New Mexico 87545, USA}
\author{R.~Mammei}                  
  \affiliation{Department of Physics, University of Winnipeg, Winnipeg, MB R3B 2E9, Canada}
\author{J.~W.~Martin}                    \affiliation{W.~K.~Kellogg Radiation Laboratory, California Institute of Technology, Pasadena, California 91125, USA}
                                         \affiliation{Department of Physics, University of Winnipeg, Winnipeg, MB R3B 2E9, Canada}
\author{D.~Melconian}                    \affiliation{Cyclotron Institute, Texas A\&M University, College Station, Texas 77843, USA}
\author{M.~P.~Mendenhall}                \thanks{Currently at Physical and Life Sciences Directorate, Lawrence Livermore National Laboratory, Livermore, CA 94550, USA}
                                         \affiliation{W.~K.~Kellogg Radiation Laboratory, California Institute of Technology, Pasadena, California 91125, USA}
\author{S.~D.~Moore}                     \affiliation{Department of Physics, North Carolina State University, Raleigh, North Carolina 27695, USA}
\author{C.~L.~Morris}                    \affiliation{Los Alamos National Laboratory, Los Alamos, New Mexico 87545, USA}
\author{S.~Nepal}                        \affiliation{Department of Physics and Astronomy, University of Kentucky, Lexington, Kentucky 40506, USA}
\author{N.~Nouri}                        \thanks{Currently at Department of Pathology, Yale University School of Medicine, New Haven, Connecticut 06510, USA}\affiliation{Department of Physics and Astronomy, University of Kentucky, Lexington, Kentucky 40506, USA}
\author{R.~W.~Pattie, Jr.}
 \thanks{Currently at Los Alamos National Lab, Los Alamos, NM 87545, USA}
\affiliation{Department of Physics, North Carolina State University, Raleigh, North Carolina 27695, USA}
                                         \affiliation{Triangle Universities Nuclear Laboratory, Durham, North Carolina 27708, USA}
\author{A.~P\'{e}rez Galv\'{a}n}         \thanks{Currently at Vertex Pharmaceuticals, 11010 Torreyana Rd., San Diego, CA 92121, USA}
                                         \affiliation{W.~K.~Kellogg Radiation Laboratory, California Institute of Technology, Pasadena, California 91125, USA}
\author{D.~G.~Phillips~II}               \affiliation{Department of Physics, North Carolina State University, Raleigh, North Carolina 27695, USA}
\author{R.~Picker}                       \thanks{Currently at TRIUMF, Vancouver, BC V6T 2A3 Canada}\affiliation{W.~K.~Kellogg Radiation Laboratory, California Institute of Technology, Pasadena, California 91125, USA}
\author{M.~L.~Pitt}                      \affiliation{Department of Physics, Virginia Tech, Blacksburg, Virginia 24061, USA}
\author{B.~Plaster}                      \affiliation{Department of Physics and Astronomy, University of Kentucky, Lexington, Kentucky 40506, USA}
\author{J.~C.~Ramsey}                    \affiliation{Los Alamos National Laboratory, Los Alamos, New Mexico 87545, USA}
\author{R.~Rios}                         \affiliation{Los Alamos National Laboratory, Los Alamos, New Mexico 87545, USA}
                                         \affiliation{Department of Physics, Idaho State University, Pocatello, Idaho 83209, USA}
\author{D.~J.~Salvat}                    \affiliation{Department of Physics and Center for Experimental Nuclear Physics and Astrophysics, University of Washington, Seattle, Washington 98195, USA}
\author{A.~Saunders}                     \affiliation{Los Alamos National Laboratory, Los Alamos, New Mexico 87545, USA}
\author{W.~Sondheim}                     \affiliation{Los Alamos National Laboratory, Los Alamos, New Mexico 87545, USA}
\author{S.~Sjue}                         \affiliation{Los Alamos National Laboratory, Los Alamos, New Mexico 87545, USA}
\author{S.~Slutsky}                      \affiliation{W.~K.~Kellogg Radiation Laboratory, California Institute of Technology, Pasadena, California 91125, USA}
\author{C.~Swank}                        \affiliation{W.~K.~Kellogg Radiation Laboratory, California Institute of Technology, Pasadena, California 91125, USA}
\author{G. Swift}                        \affiliation{Triangle Universities Nuclear Laboratory, Durham, North Carolina 27708, USA}
\author{E.~Tatar}                        \affiliation{Department of Physics, Idaho State University, Pocatello, Idaho 83209, USA}
\author{R.~B.~Vogelaar}                  \affiliation{Department of Physics, Virginia Tech, Blacksburg, Virginia 24061, USA}
\author{B.~VornDick}                     \affiliation{Department of Physics, North Carolina State University, Raleigh, North Carolina 27695, USA}
\author{Z.~Wang}                         \affiliation{Los Alamos National Laboratory, Los Alamos, New Mexico 87545, USA}
\author{W.~Wei}
 \affiliation{W.~K.~Kellogg Radiation Laboratory, California Institute of Technology, Pasadena, California 91125, USA}
\author{J.~Wexler}                       \affiliation{Department of Physics, North Carolina State University, Raleigh, North Carolina 27695, USA}
\author{T.~Womack}                       \affiliation{Los Alamos National Laboratory, Los Alamos, New Mexico 87545, USA}
\author{C.~Wrede}                        \affiliation{Department of Physics and Center for Experimental Nuclear Physics and Astrophysics, University of Washington, Seattle, Washington 98195, USA}
                                         \affiliation{Department of Physics and Astronomy and National Superconducting Cyclotron Laboratory, Michigan State University, East Lansing, Michigan 48824, USA}
\author{A.~R.~Young}                     \affiliation{Department of Physics, North Carolina State University, Raleigh, North Carolina 27695, USA}
                                         \affiliation{Triangle Universities Nuclear Laboratory, Durham, North Carolina 27708, USA}
\author{B.~A.~Zeck}                      \affiliation{Department of Physics, North Carolina State University, Raleigh, North Carolina 27695, USA}

\collaboration{UCNA Collaboration}
\noaffiliation

\date{\today}

\begin{abstract}
It has been proposed recently that a previously unobserved neutron decay branch to a dark matter particle ($\chi$) could account for the discrepancy in the neutron lifetime observed in experiments that use two different measurement techniques. One of the possible final states discussed includes a single $\chi$ along with an $e^{+}e^{-}$ pair. We use data from the UCNA (Ultracold Neutron Asymmetry) experiment to set limits on this decay channel. Coincident electron-like events are detected with $\sim 4\pi$ acceptance using a pair of detectors that observe a volume of stored Ultracold Neutrons (UCNs). The summed kinetic energy ($E_{e^{+}e^{-}}$) from such events is used to set limits, as a function of the $\chi$ mass, on the branching fraction for this decay channel. 
For $\chi$ masses consistent with resolving the neutron lifetime discrepancy, 
we exclude this as the dominant dark matter decay channel at $\gg~5\sigma$ level for $100~\text{keV} < E_{e^{+}e^{-}} < 644~\text{keV}$. 
If the $\chi+e^{+}e^{-}$ final state is not the only one, we set limits on its branching fraction of $< 10^{-4}$ for the above $E_{e^{+}e^{-}}$ range at $> 90\%$ confidence level.  
\end{abstract}

\maketitle

Precise measurements of the neutron lifetime using two different techniques yield values that disagree at the 4$\sigma$ level \cite{Greene:2016,RevModPhys.83.1173,Young2014mxa}. In one technique, decay protons are collected and counted for a fixed length of a neutron beam and compared to the number of neutrons in that beam.  In the second, the number of neutrons remaining in a storage vessel are counted after different storage times. These storage vessel experiments are termed material ``bottle'' experiments and were recently confirmed by a magneto-gravitational trap bottle experiment \cite{Pattie:2017vsj}. A recent theoretical explanation for this lifetime difference \cite{Fornal:2018eol} suggests that a neutron decaying to a proton is not the only possible decay mode, but that a decay to a new dark matter particle is also possible. If the branching ratio of the dark matter decay to neutron decay is $\sim$1\%, this would account for the lifetime anomaly.

This proposal has generated a number of new studies including constraints from neutron star formation \cite{McKeen:2018xwc,Baym:2018ljz,Motta:2018rxp,Cline:2018ami}, constraints from precision neutron and nuclear $\beta$-decay studies \cite{Czarnecki:2018okw}, a new direct search for the dark matter decay $n\rightarrow \chi+\gamma$ \cite{Tang:2018eln}, and a proposal for a future search using nuclear beta decay \cite{Pfutzner:2018ieu}. 

In this work we use the latest data from the Ultracold Neutron Asymmetry (UCNA) experiment \cite{Plaster2012,Mendenhall2013,Brown:2017mhw,Hickerson:2017fzz,Holley2012} to put direct constraints on one of the proposed dark matter decay channels: $n\rightarrow \chi+e^{+}e^{-}$. In this decay mode, the $e^{+}e^{-}$ sum energy is approximately equal to the entire mass difference between the neutron and the $\chi$. At a $\sim$1\% branching ratio, this would yield a clear peak in the UCNA detector energy spectrum.

\begin{figure}
\includegraphics[width=\columnwidth]{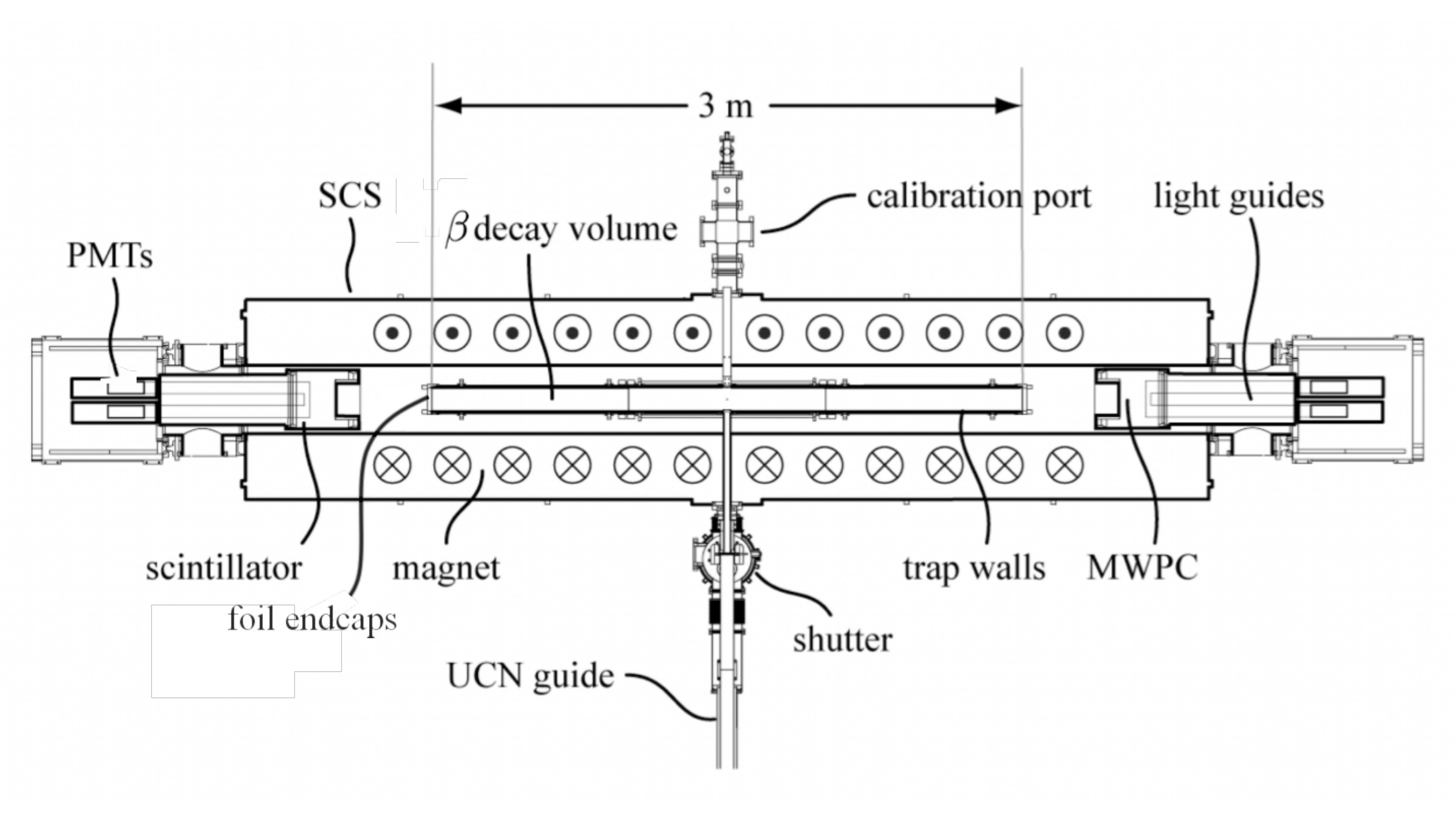}
\caption{Schematic diagram of the UCNA spectrometer.}
\label{fig:scs-section-detail}
\end{figure}

The UCNA experiment, located at the Los Alamos Neutron Science Center (LANSCE), has been described in \cite{Plaster2012}. A schematic diagram of the UCNA apparatus is shown in Fig. \ref{fig:scs-section-detail}, and a brief summary of the apparatus relevant to the present analysis is provided here for context.

Neutrons are produced from a tungsten spallation target \cite{Saunders2003,Morris2002,Saunders2013}, cooled down to UCN energies (kinetic energy $<$ 350~neV) and transported to the spectrometer. Here the UCNs are contained within a 3~m long decay trap in the 5~m long superconducting spectrometer (SCS). In the SCS, a 1~T magnetic field directs the decay electrons toward two detectors located on either end \cite{PlasterSCS2008}, which will hereafter be called the `East' and `West' detectors. The neutrons are polarized \cite{Holley2012} when they arrive in the SCS and hence are either aligned or anti-aligned with the magnetic field. However, the present analysis averages over the polarization, and thus polarization effects are not considered.

Each electron detector consists of a multiwire proportional chamber (MWPC) \cite{ItoMWPC2007,Morris2009} followed by a 3.5~mm thick plastic scintillator. The MWPC provides position reconstruction and ``backscattering'' identification, i.e. electrons that scatter and produce a signal in both detectors. This position reconstruction allows us to define a fiducial volume that is free from events scattering off the decay trap walls. The main plastic scintillator provides timing and energy reconstruction. The timing information is based on CAEN V775 time-to-digital converters (TDCs). For background suppression of cosmic rays, several veto detectors are used: scintillator paddles and argon/ethane drift tubes \cite{Rios2011105} placed above and on the side of the detectors, as well as 15~cm diameter, 25~mm thick scintillators placed directly behind the electron detectors. 

For the present analysis, only data taken during the most recent, 2012-2013, UCNA run is used, since timing information is crucial for background suppression in this work and the TDCs were operating most reliably during this period. Older datasets show small systematic timing drifts, which had minimal impact on the $\beta$-decay asymmetry analysis \cite{Brown:2017mhw}, but would introduce significant additional background in this work. Thus, a total of approximately 14.55 million neutron decays are considered for the present analysis, after applying all cuts and correcting for conventional $\beta$-decay detection efficiency \cite{Mendenhall2014}.

This analysis focuses on electron-like pair events that produce a short time-coincidence between both ends of the SCS, since this provides a particularly clean signature. In standard neutron $\beta$-decay analysis, coincidence events occur due to a small fraction ($\sim 3.8\%$) of cases where the single decay electron is scattered in the first detector after depositing a fraction of its energy and then traverses the spectrometer to deposit the remainder of its energy in the other detector - the aforementioned ``backscattering'' events. However, for these events, there is a minimum time required for the electron to traverse the spectrometer: the scintillator-to-scintillator distance is 4.4~m and thus a single, maximum energy $\beta$-decay electron pointing directly towards one detector cannot trigger both scintillators with a time difference $<$ 16~ns. Any events that trigger both ends within this minimum crossing time are candidate dark matter decays to $e^{+}e^{-}$ within the neutron decay volume. Such dark matter decay events will generally not be detected simultaneously because of differences in energies, pitch-angles relative to the 1~T magnetic field in the SCS, and distances to each detector since UCNs populate the 3~m long decay trap nearly uniformly \cite{PlasterSCS2008}.

\begin{figure}
\includegraphics[width=\linewidth]{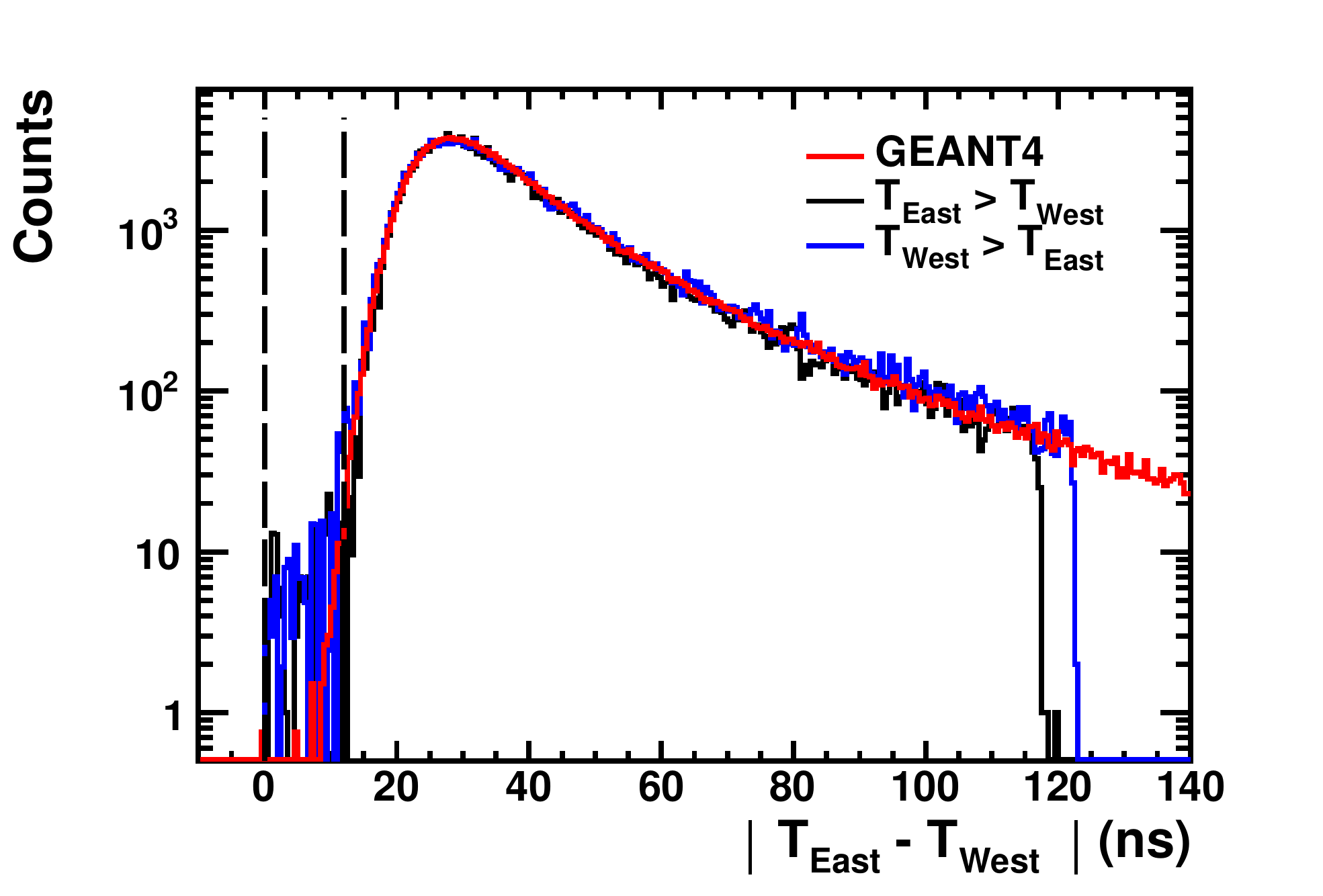}

\caption{Background-subtracted and calibrated timing distributions of events that trigger both East and West detectors, separated according to which detector triggered first. Overlaid is a GEANT4 simulation of the time-of-flight for all backscattering $\beta$-decay events that trigger both East and West detectors. The simulation includes a timing resolution of 3~ns which best matches the data. The time cutoff near 120~ns for the data is an instrumental effect. Dashed lines indicate the chosen analysis time-window (see text).} 
\label{fig:fig-timing-spectra}
\end{figure}

For coincidence signals, the relative time between detector events is formed separately for the East and West detectors with the first trigger arriving at the electronics producing a common stop signal for both sides. The relative cable delay between each side is determined from the background events spectrum. Background is measured in dedicated background runs by acquiring data with the UCNs blocked by a gate value approximately 7~m upstream of the spectrometer. These background events are dominated by high energy photons that produce electrons, via Compton scattering, in one detector that then travel toward the opposite end, producing a signal in both. These events, which typically have a higher energy than neutron $\beta$-decay, have a sharp turn-on time difference of 15.5~ns.

For each detector, the background-subtracted distributions of time differences for all coincidence events are shown in Fig. \ref{fig:fig-timing-spectra}. The large signal with a peak at $\sim 25$~ns is caused by backscattering electrons from neutron $\beta$-decay events. This is explicitly shown by the overlaid, normalized, resolution-corrected, GEANT4 simulation of neutron $\beta$-decay backscatters which includes our spectrometer and detector effects \cite{Mendenhall2014,Brown:2017mhw}. Thus, the timing region for dark matter neutron decay is where the time difference is $<$ 16~ns. Because of the finite, 3~ns timing resolution of the detectors, backscattering events can begin to dominate the signal at times earlier than the expected 16~ns discussed above. Based on studies with a variable time-cut, and a GEANT4 simulation of the time spectrum of conventional backscatter events from neutron $\beta$-decay, a time-window of 0 to 12~ns is chosen for candidate dark decay $e^{+}e^{-}$ events.

Along with the timing information, the summed kinetic energy of the leptons, $E_{e^+e^-}$, further improves signal-to-background of the dark decay search. Because $E_{e^+e^-}$ is nearly mono-energetic (since the kinetic energy of the $\chi$ particle satisfies $E_{\chi} \ll E_{e^{+} e^{-}} \ll m_\chi$) a narrow energy window can be used. This summed energy is given by $E_{e^+e^-}\simeq m_n-m_{\chi}-2m_e$, where $m_{n}$, $m_{\chi}$, and $m_{e}$ are the masses of the neutron, the dark sector particle $\chi$, and the electron (or positron), respectively. Of course, the energy resolution of the detectors, which is measured \cite{Mendenhall2013,Hickerson:2017fzz,Brown:2017mhw} to be 
$\Delta E/E = 0.05 \times \sqrt{E/1~\text{MeV}}$, broadens this signal and is accounted for in this analysis. Furthermore, while the detector energy response for incident electrons has been carefully calibrated \cite{Brown:2017mhw}, the response for incident positrons is determined from a GEANT4 simulation using our full detector geometry.\footnote{The GEANT4 framework is described in more detail in \cite{Geant4}.} This simulation indicates that the fraction of positrons that deposit only their full kinetic energy in the scintillator is approximately 15\% smaller than that of electrons over the energy range of interest, primarily due to annihilation. This reduction is accounted for in the calculation of the final limits produced from the data.

\begin{figure}
\subfloat[]{\label{fig:bg-erecon-full}
  \includegraphics[trim={0 0 0 1.35cm},clip,width=\columnwidth,height=0.6\columnwidth]{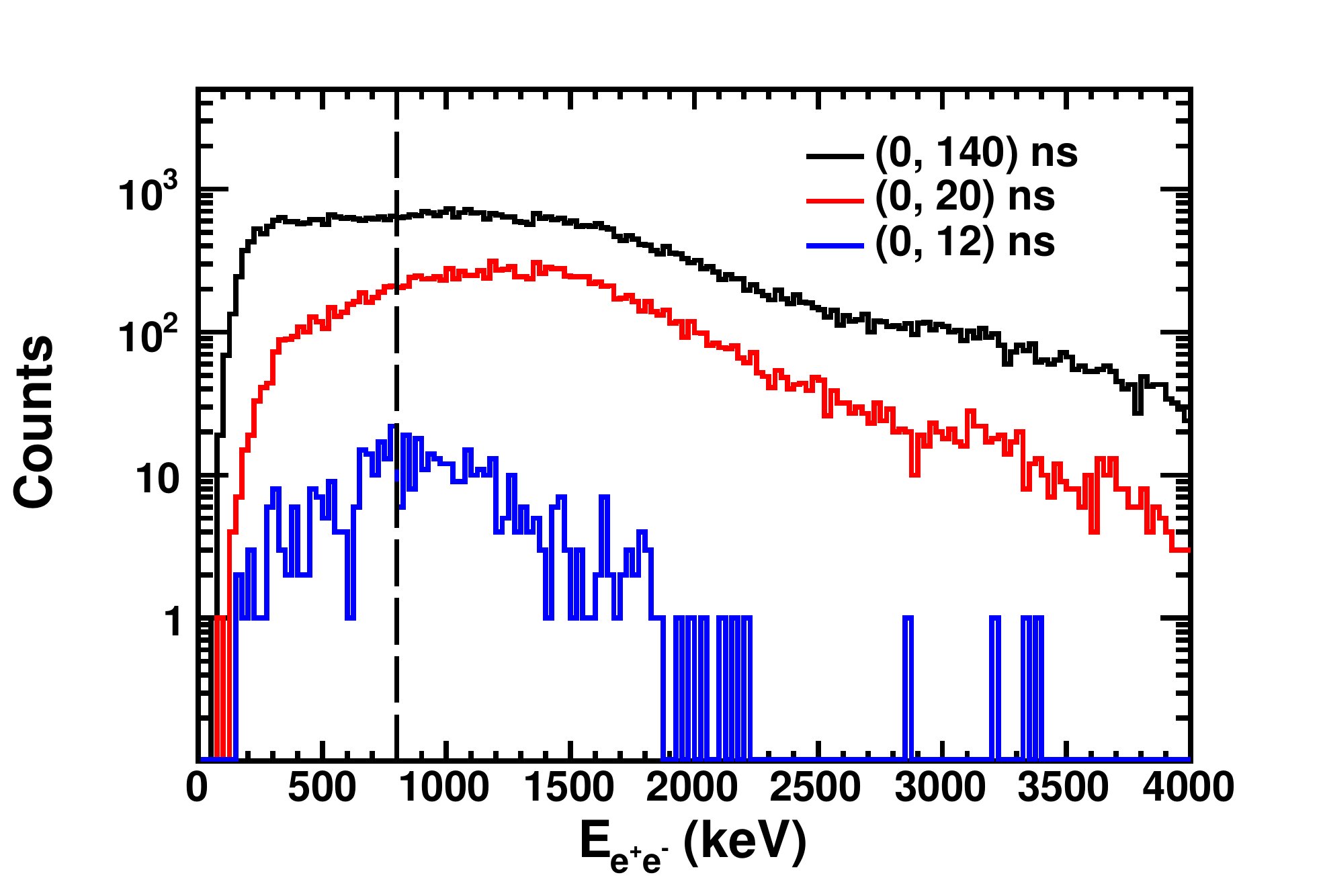}
}
\vskip -.53in
\subfloat[]{\label{fig:fg-erecon-full}
  \includegraphics[trim={0 0 0 1.50cm},clip,width=\columnwidth,height=0.6\columnwidth]{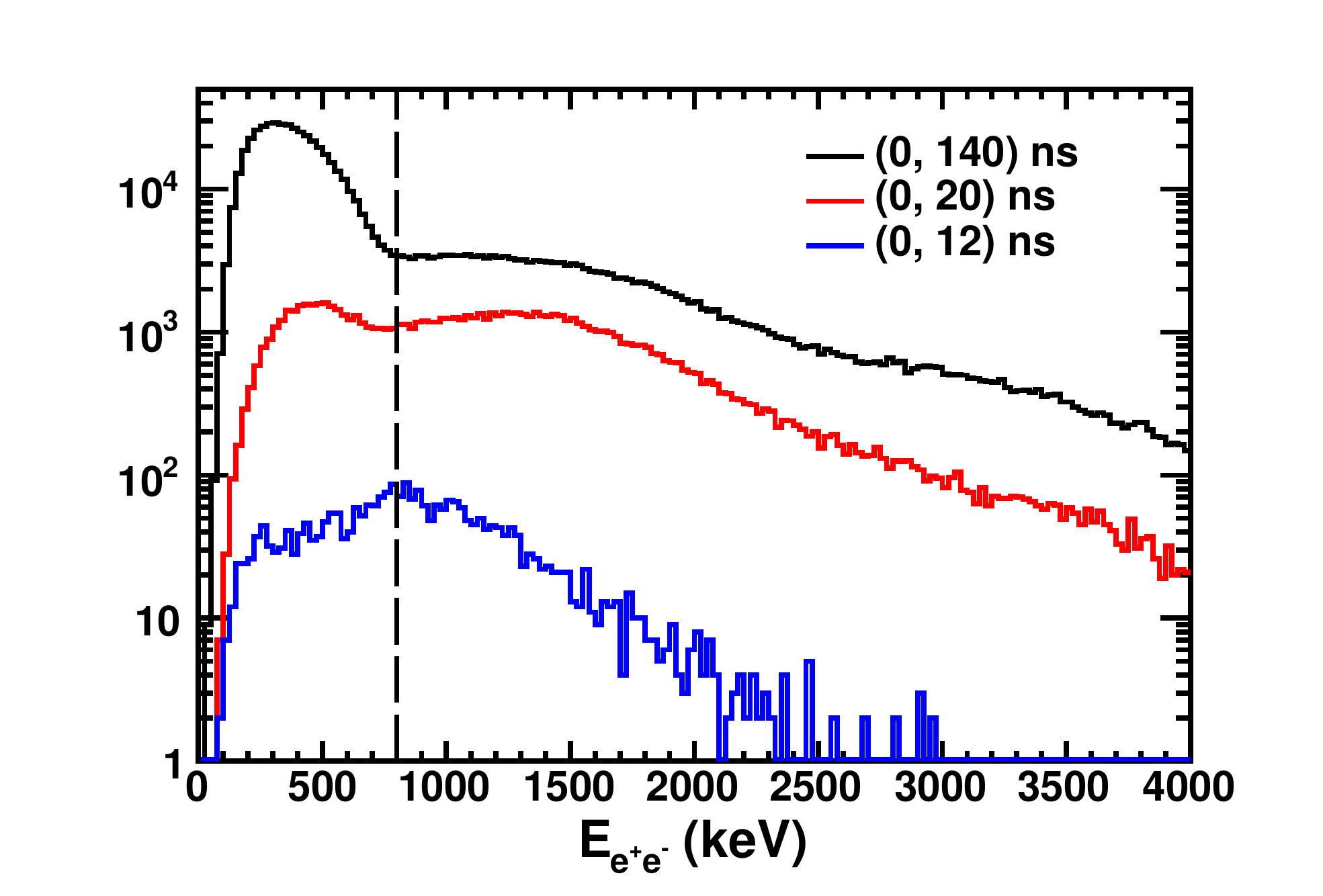}
}

\caption{\label{fig:erecon-spectra-full} Energy spectrum of (top) background and (bottom) foreground runs, for three separate time-windows. Clear structure of a neutron $\beta$-decay backscattering peak at 300~keV is visible for time-windows $>$ 12~ns in the foreground runs. Dashed lines at 0~keV, 800~keV indicate the energy region of interest used for the present analysis.}
\end{figure}

The energy spectra resulting from applying coincidence-timing cuts to the UCNA runs, foreground and background, are shown in Fig. \ref{fig:erecon-spectra-full} over the full energy range of the detectors.
The spectra are shown for several cuts on relative time between triggers to show the impact of the timing cuts on reducing the potential background due to ordinary $\beta$-decay in the dark decay region of interest.  Note that the foreground shows a predominantly neutron-induced signal only for event windows with time differences nominally greater than 16 ns. This appears as a peak in energy at 300~keV with an endpoint near 800~keV which is the signature of neutron $\beta$-decay backscatter events, as expected.

\begin{figure}

\subfloat[]{\label{fig:bgSub-erecon}
  \includegraphics[trim={0 0 0 1.0cm},clip,width=\columnwidth,height=0.6\columnwidth]
  {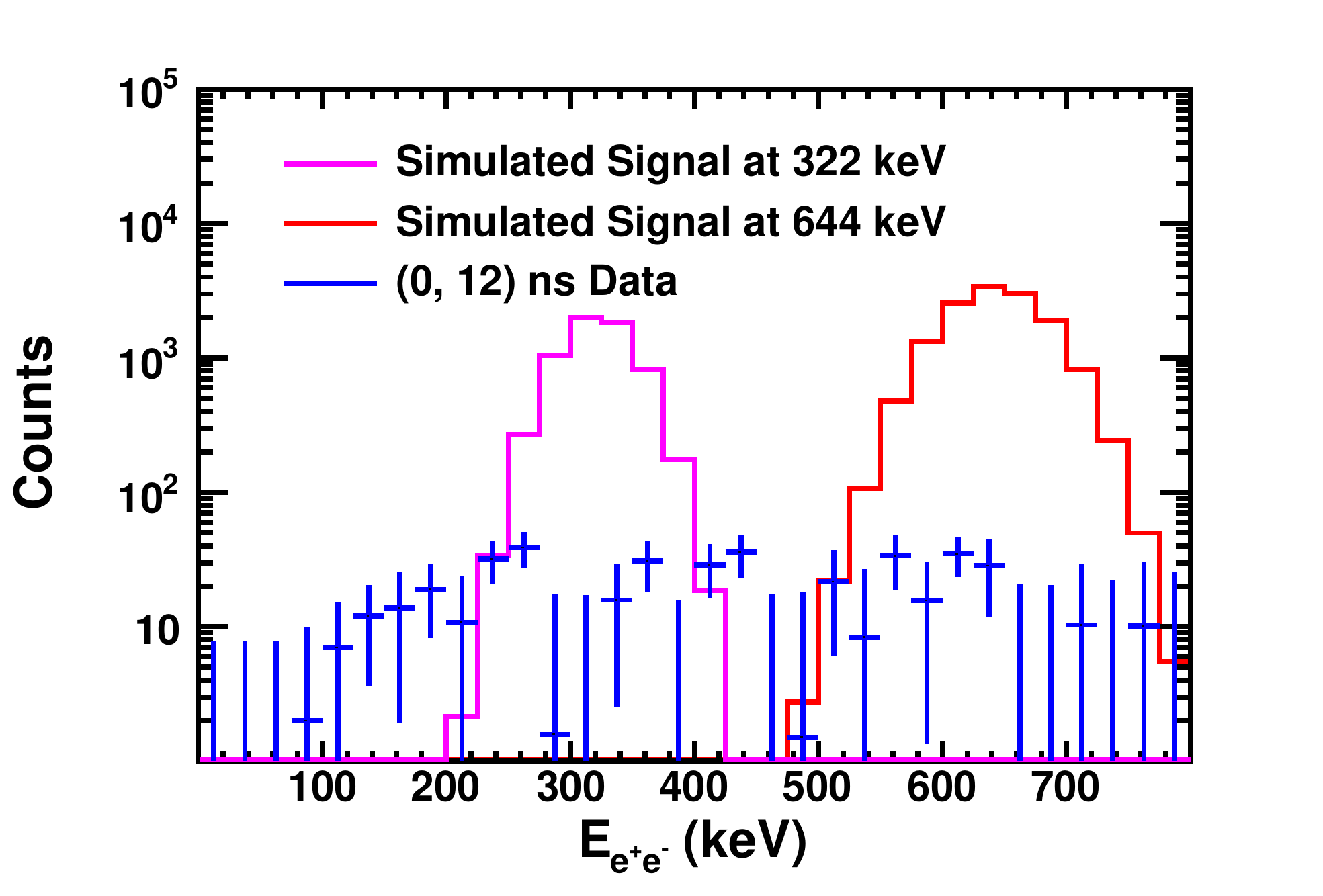}
}
\vskip -.52in
\subfloat[]{\label{fig:acceptance}
  \includegraphics[trim={0 0 0 1.5cm},clip,width=\columnwidth,height=0.6\columnwidth]
  {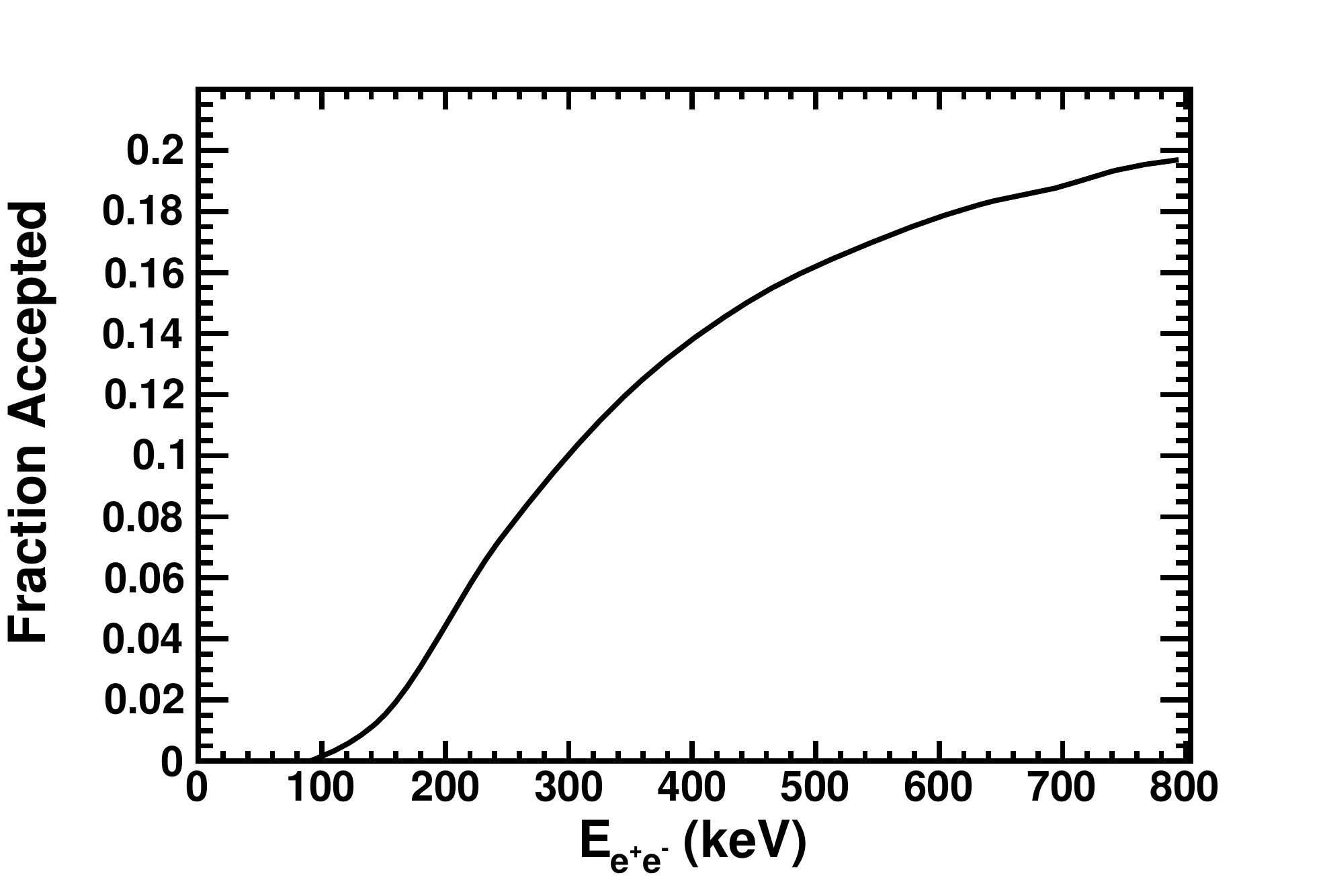}
}

\caption{\label{fig:erecon-valid} (Top) background-subtracted $e^+e^-$ pair kinetic energy spectra for events in the analysis time-window. For comparison, simulated positive dark matter decay signals at 322~keV, 644~keV are overlaid. (Bottom) total $e^+e^-$ pair acceptance (see text) as a function of summed kinetic energy.}

\end{figure}

Neutron decay events can be studied once events from background runs are subtracted. However the data for the UCNA experiment was taken with a large signal-to-noise ratio ($\frac{S}{N} \gg 1$) which was optimized for the measurement of the $A$ asymmetry parameter \cite{Mendenhall2013}. This corresponds to a relative live-time of 5:1 for foreground to background runs. For this analysis, attempting to set a limit on potential dark-matter $\chi$ decays (with, possibly, $\frac{S}{N} < 1$) means that a background-subtracted analysis will have error bars dominated by the background. Poisson statistics are used when the counts per bin are low, which become conventional Gaussian statistics when the counts become higher. The background-subtracted spectrum of $e^{+}e^{-}$ summed kinetic energy is shown in the top panel of Fig. \ref{fig:erecon-valid}. The proposed $e^{+}e^{-}$ decay channel in \cite{Fornal:2018eol} has a valid sum total energy range for the $e^{+}e^{-}$ pair of between $2 m_e$ and 1.665~MeV, which translates to a sum kinetic energy range for $E_{e^{+}e^{-}}$ of 0 to 644~keV. Two simulated positive $n\rightarrow\chi+e^{+}e^{-}$ signals with a 1\% branching ratio (corrected for efficiencies, discussed below) are also shown in this figure, corresponding to two possible values of $m_{\chi}$. This is the scale of branching ratio that would be needed to explain the neutron lifetime anomaly in the case of exclusive decay to $e^{+}e^{-}$.

In order to place appropriate limits on the rate of the dark matter $e^{+}e^{-}$ decay, the number of observed events must be scaled for kinematic efficiency, particle detection efficiency, and analysis cut efficiency. 

The kinematic efficiency accounts for decays that produce $e^{+}e^{-}$ pairs directed toward only one of the detectors. These events must be rejected because they do not have the relative time information that dramatically reduces the background signal from the standard decay: $n\rightarrow p e^{-} \bar{\nu}$. The kinematic efficiency is calculated via Monte Carlo simulation using a uniform population of the available phase space for the three-particle decay: $\chi +e^+e^-$. Thus, events are selected where the $e^{+}e^{-}$ are directed to separate detectors, which corresponds to $\sim 60\%$ of the decays at all values of $m_\chi$. 

At electron kinetic energies above 200~keV, the particle detection efficiency in each detector is essentially 100\%. This drops off rapidly below 100~keV because of energy loss in the decay trap windows and the MWPCs, reaching 50\% at 75~keV. Since we require a detection in both the East and West detectors, this efficiency must be applied twice. In addition, there is a reduced efficiency for capturing the full energy of the positron compared to the electron (discussed above), which is also applied.  

Lastly, there is a timing-window cut efficiency. A short timing-window cut reduces efficiency while a long timing-window cut increases background due to backscatter contamination. Monte Carlo simulations of decay products produced uniformly within the 3~m decay trap are used to estimate this efficiency, for various time-windows. Fig. \ref{fig:erecon-spectra-full} shows the data's reconstructed energy spectra for some representative time-windows - several more were generated and examined for backscatter contamination to determine a final time-window. It is evident from our analysis, and shown in this figure, that a time-window endpoint of 12~ns avoids the backscatter contamination seen in the larger time-windows. Considering these effects, a timing-window cut of 0-12~ns is used, illustrated in Fig. \ref{fig:fig-timing-spectra}. The efficiency of this chosen timing-window cut ranges from $\sim$ 20\% to 40\% for the full range of $E_{e^+e^-}$.

The total acceptance as a function of $E_{e^+e^-}$, which includes the detection fraction based on the decay kinematics, the trigger efficiency with the reduced positron full-energy deposition efficiency discussed above, and the timing window, can be seen in the bottom panel of Fig. \ref{fig:erecon-valid}. 

Final exclusion confidence limits are determined using the background-subtracted dataset binned into discrete energy bins with width comparable to the energy resolution, and checked for bin aliasing. Since there is no evidence of a peak structure in the data, a 1$\sigma$ detection limit is then determined from the 1$\sigma$ uncertainty in each of these bins (assuming that a hypothesized signal has fluctuated down to the background level).  While this analysis attempts to set a limit on the existence of a peak at a single energy, it is not known beforehand at what energy this peak should occur. Thus, since we are searching over a range in energies, fluctuations at other energies must be considered. This is usually termed the ``look-elsewhere effect'' - the probability that a statistically significant fluctuation will occur given enough samples \cite{lyons2008}. 

\begin{figure}
\includegraphics[width=\linewidth]{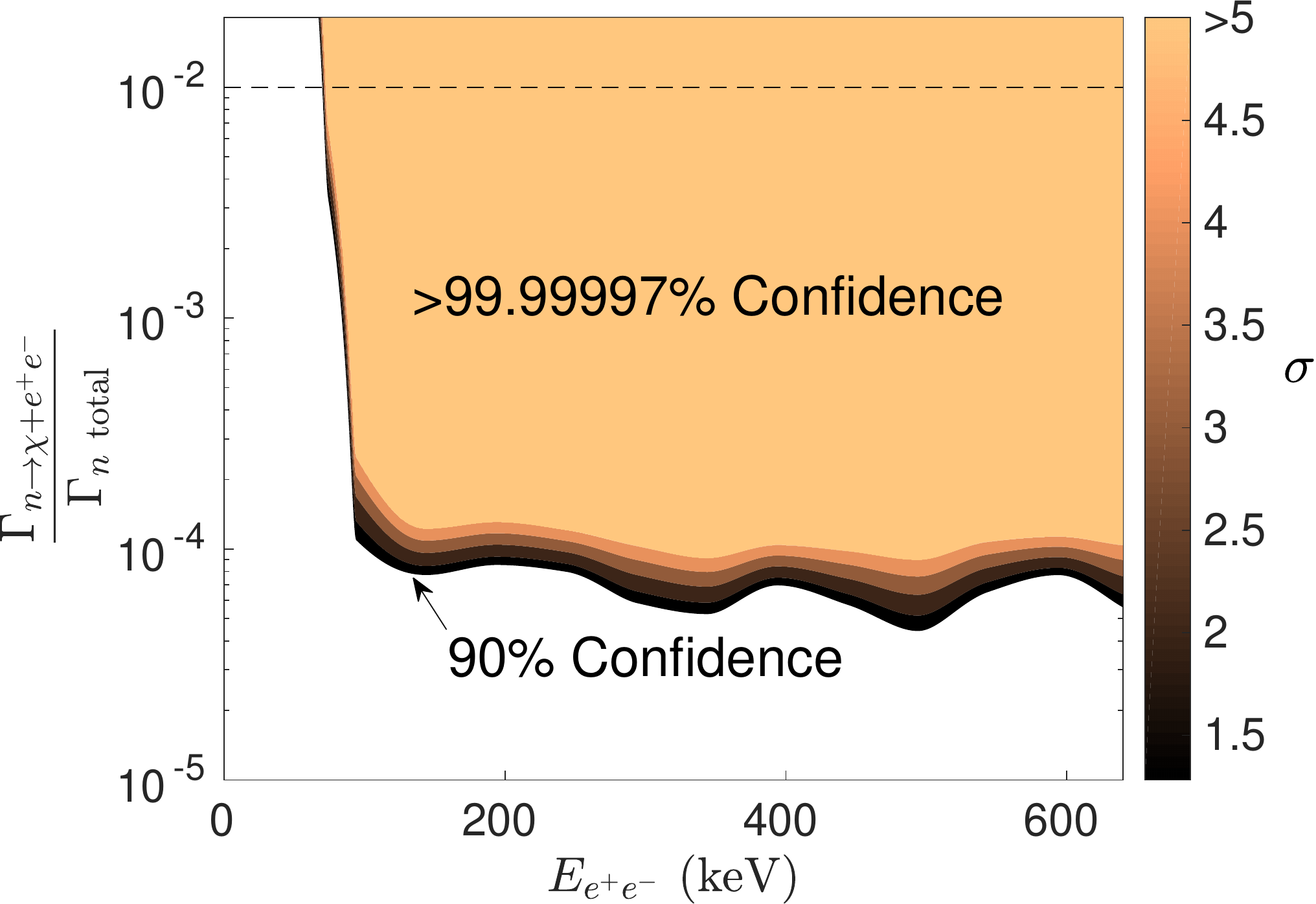}
\vspace{-.3in}
\caption{Confidence limits on the branching ratio of the neutron dark decay channel, as a function of the kinetic energy of the produced $e^{+} e^{-}$ pair. This is directly related to the proposed $\chi$ mass by $m_{\chi} = m_{n} - 2m_{e} - E_{e^{+}e^{-}}$, which has a range of $937.900~\text{MeV} < m_{\chi} < 938.543~\text{MeV}$. A branching ratio of $10^{-2}$, which would be required to explain the neutron lifetime anomaly if $n\rightarrow \chi + e^{+}e^{-}$ were the only allowed final state, is shown by the dashed line.}
\label{fig:confidence-contours}
\end{figure}

This look-elsewhere effect was accounted for numerically. First, a statistical test was constructed,  
\begin{align}
\xi = \sum_i \frac{N_i-\mu_i}{\sigma_i} \mathrm{~for~} N_i>\mu_i
\end{align}
where $N_i$ is a normally-distributed random variable for bin $i$ with mean $\mu_i$ and standard deviation $\sigma_i$, and both $\mu_i$ and $\sigma_i$ are given by the data. The test was then computed for the distribution of events for the final energy bins. A single bin probability distribution function (PDF) for $\xi$ is then compared to a $\xi$ PDF for the final energy bins using a large number of Monte Carlo samples. The ratio of confidence levels for these two tests provides the ``look-elsewhere'' correction factor. 
These counts for a given confidence level are then corrected for the acceptance discussed above as well as the effects of the finite energy resolution to provide a corresponding limit on the possible number of dark matter decays. Branching ratio limits are then produced by dividing by the total number of $\beta$-decays ($=$ 14.55 million), which is corrected for the single electron detection efficiency \cite{Brown:2017mhw,Hickerson:2017fzz,Mendenhall2014}. These limits are shown in Fig. \ref{fig:confidence-contours}, where the dark matter $\chi + e^{+}e^{-}$ final state is ruled out at $\gg 5\sigma$ for $E_{e^+e^-}> 100~\text{keV}$ at the $1\%$ branching ratio required to resolve the lifetime anomaly. In addition, as discussed in Ref. \cite{Fornal:2018eol}, the $\chi + e^{+}e^{-}$ final state could be suppressed compared to the $\chi + \gamma$ final state. If this is the case, we set a limit at the $> 90\%$ confidence level on the branching ratio to the dark matter decay of $\frac{\Gamma_{n\rightarrow\chi + e^{+}e^{-}}}{\Gamma_{n}} < 10^{-4}$ down to $E_{e^{+}e^{-}} = 100$~keV.

In summary, we have used $\sim 1.5\times 10^{7}$ free neutron decays from the UCNA measurement to perform a direct search for neutron decay to a dark particle: $n \rightarrow \chi +e^{+}e^{-}$. Using timing information with a two-detector trigger, background from normal neutron decay can be dramatically suppressed.
We find that if $\chi + e^{+}e^{-}$ were the dominant dark matter decay channel with branching ratio required to resolve the neutron lifetime discrepancy, it is ruled out at $\gg 5\sigma$ level for all $\chi$ masses corresponding to $100~\text{keV} < E_{e^+e^-} < 644~\text{keV}$. 

This work is supported in part by the US Department
of Energy, Office of Nuclear Physics (DE-FG02-08ER41557, DE-SC0014622, DE-FG02-97ER41042) and the
National Science Foundation
(1002814,
1005233, 
1205977, 
1306997
1307426, 
1506459, and
1615153).
We gratefully acknowledge the support of the LDRD program (20110043DR), the AOT division of the Los Alamos National
Laboratory, and helpful discussions with B. Grinstein.

\bibliography{references.bib}

\begin{thebibliography}{27}%
\makeatletter
\providecommand \@ifxundefined [1]{%
 \@ifx{#1\undefined}
}%
\providecommand \@ifnum [1]{%
 \ifnum #1\expandafter \@firstoftwo
 \else \expandafter \@secondoftwo
 \fi
}%
\providecommand \@ifx [1]{%
 \ifx #1\expandafter \@firstoftwo
 \else \expandafter \@secondoftwo
 \fi
}%
\providecommand \natexlab [1]{#1}%
\providecommand \enquote  [1]{``#1''}%
\providecommand \bibnamefont  [1]{#1}%
\providecommand \bibfnamefont [1]{#1}%
\providecommand \citenamefont [1]{#1}%
\providecommand \href@noop [0]{\@secondoftwo}%
\providecommand \href [0]{\begingroup \@sanitize@url \@href}%
\providecommand \@href[1]{\@@startlink{#1}\@@href}%
\providecommand \@@href[1]{\endgroup#1\@@endlink}%
\providecommand \@sanitize@url [0]{\catcode `\\12\catcode `\$12\catcode
  `\&12\catcode `\#12\catcode `\^12\catcode `\_12\catcode `\%12\relax}%
\providecommand \@@startlink[1]{}%
\providecommand \@@endlink[0]{}%
\providecommand \url  [0]{\begingroup\@sanitize@url \@url }%
\providecommand \@url [1]{\endgroup\@href {#1}{\urlprefix }}%
\providecommand \urlprefix  [0]{URL }%
\providecommand \Eprint [0]{\href }%
\providecommand \doibase [0]{http://dx.doi.org/}%
\providecommand \selectlanguage [0]{\@gobble}%
\providecommand \bibinfo  [0]{\@secondoftwo}%
\providecommand \bibfield  [0]{\@secondoftwo}%
\providecommand \translation [1]{[#1]}%
\providecommand \BibitemOpen [0]{}%
\providecommand \bibitemStop [0]{}%
\providecommand \bibitemNoStop [0]{.\EOS\space}%
\providecommand \EOS [0]{\spacefactor3000\relax}%
\providecommand \BibitemShut  [1]{\csname bibitem#1\endcsname}%
\let\auto@bib@innerbib\@empty
\bibitem [{\citenamefont {Greene}\ and\ \citenamefont
  {Geltenbort}(2016)}]{Greene:2016}%
  \BibitemOpen
  \bibfield  {author} {\bibinfo {author} {\bibfnamefont {G.~L.}\ \bibnamefont
  {Greene}}\ and\ \bibinfo {author} {\bibfnamefont {P.}~\bibnamefont
  {Geltenbort}},\ }\href@noop {} {\bibfield  {journal} {\bibinfo  {journal}
  {Sci. Am.}\ }\textbf {\bibinfo {volume} {314}},\ \bibinfo {pages} {37}
  (\bibinfo {year} {2016})}\BibitemShut {NoStop}%
\bibitem [{\citenamefont {Wietfeldt}\ and\ \citenamefont
  {Greene}(2011)}]{RevModPhys.83.1173}%
  \BibitemOpen
  \bibfield  {author} {\bibinfo {author} {\bibfnamefont {F.~E.}\ \bibnamefont
  {Wietfeldt}}\ and\ \bibinfo {author} {\bibfnamefont {G.~L.}\ \bibnamefont
  {Greene}},\ }\href {\doibase 10.1103/RevModPhys.83.1173} {\bibfield
  {journal} {\bibinfo  {journal} {Rev. Mod. Phys.}\ }\textbf {\bibinfo {volume}
  {83}},\ \bibinfo {pages} {1173} (\bibinfo {year} {2011})}\BibitemShut
  {NoStop}%
\bibitem [{\citenamefont {Young}\ \emph {et~al.}(2014)\citenamefont {Young}
  \emph {et~al.}}]{Young2014mxa}%
  \BibitemOpen
  \bibfield  {author} {\bibinfo {author} {\bibfnamefont {A.~R.}\ \bibnamefont
  {Young}} \emph {et~al.},\ }\href {\doibase 10.1088/0954-3899/41/11/114007}
  {\bibfield  {journal} {\bibinfo  {journal} {J. Phys.}\ }\textbf {\bibinfo
  {volume} {G41}},\ \bibinfo {pages} {114007} (\bibinfo {year}
  {2014})}\BibitemShut {NoStop}%
\bibitem [{\citenamefont {Pattie}\ \emph {et~al.}(2017)\citenamefont {Pattie}
  \emph {et~al.}}]{Pattie:2017vsj}%
  \BibitemOpen
  \bibfield  {author} {\bibinfo {author} {\bibfnamefont {R.~W.}\ \bibnamefont
  {Pattie}, \bibfnamefont {Jr.}} \emph {et~al.},\ }\href@noop {} {\  (\bibinfo
  {year} {2017})},\ \Eprint {http://arxiv.org/abs/1707.01817} {arXiv:1707.01817
  [nucl-ex]} \BibitemShut {NoStop}%
\bibitem [{\citenamefont {Fornal}\ and\ \citenamefont
  {Grinstein}(2018)}]{Fornal:2018eol}%
  \BibitemOpen
  \bibfield  {author} {\bibinfo {author} {\bibfnamefont {B.}~\bibnamefont
  {Fornal}}\ and\ \bibinfo {author} {\bibfnamefont {B.}~\bibnamefont
  {Grinstein}},\ }\href@noop {} {\  (\bibinfo {year} {2018})},\ \Eprint
  {http://arxiv.org/abs/1801.01124} {arXiv:1801.01124 [hep-ph]} \BibitemShut
  {NoStop}%
\bibitem [{\citenamefont {McKeen}\ \emph {et~al.}(2018)\citenamefont {McKeen},
  \citenamefont {Nelson}, \citenamefont {Reddy},\ and\ \citenamefont
  {Zhou}}]{McKeen:2018xwc}%
  \BibitemOpen
  \bibfield  {author} {\bibinfo {author} {\bibfnamefont {D.}~\bibnamefont
  {McKeen}}, \bibinfo {author} {\bibfnamefont {A.~E.}\ \bibnamefont {Nelson}},
  \bibinfo {author} {\bibfnamefont {S.}~\bibnamefont {Reddy}}, \ and\ \bibinfo
  {author} {\bibfnamefont {D.}~\bibnamefont {Zhou}},\ }\href@noop {} {\
  (\bibinfo {year} {2018})},\ \Eprint {http://arxiv.org/abs/1802.08244}
  {arXiv:1802.08244 [hep-ph]} \BibitemShut {NoStop}%
\bibitem [{\citenamefont {Baym}\ \emph {et~al.}(2018)\citenamefont {Baym},
  \citenamefont {Beck}, \citenamefont {Geltenbort},\ and\ \citenamefont
  {Shelton}}]{Baym:2018ljz}%
  \BibitemOpen
  \bibfield  {author} {\bibinfo {author} {\bibfnamefont {G.}~\bibnamefont
  {Baym}}, \bibinfo {author} {\bibfnamefont {D.~H.}\ \bibnamefont {Beck}},
  \bibinfo {author} {\bibfnamefont {P.}~\bibnamefont {Geltenbort}}, \ and\
  \bibinfo {author} {\bibfnamefont {J.}~\bibnamefont {Shelton}},\ }\href@noop
  {} {\  (\bibinfo {year} {2018})},\ \Eprint {http://arxiv.org/abs/1802.08282}
  {arXiv:1802.08282 [hep-ph]} \BibitemShut {NoStop}%
\bibitem [{\citenamefont {Motta}\ \emph {et~al.}(2018)\citenamefont {Motta},
  \citenamefont {Guichon},\ and\ \citenamefont {Thomas}}]{Motta:2018rxp}%
  \BibitemOpen
  \bibfield  {author} {\bibinfo {author} {\bibfnamefont {T.~F.}\ \bibnamefont
  {Motta}}, \bibinfo {author} {\bibfnamefont {P.~A.~M.}\ \bibnamefont
  {Guichon}}, \ and\ \bibinfo {author} {\bibfnamefont {A.~W.}\ \bibnamefont
  {Thomas}},\ }\href@noop {} {\  (\bibinfo {year} {2018})},\ \Eprint
  {http://arxiv.org/abs/1802.08427} {arXiv:1802.08427 [nucl-th]} \BibitemShut
  {NoStop}%
\bibitem [{\citenamefont {Cline}\ and\ \citenamefont
  {Cornell}(2018)}]{Cline:2018ami}%
  \BibitemOpen
  \bibfield  {author} {\bibinfo {author} {\bibfnamefont {J.~M.}\ \bibnamefont
  {Cline}}\ and\ \bibinfo {author} {\bibfnamefont {J.~M.}\ \bibnamefont
  {Cornell}},\ }\href@noop {} {\  (\bibinfo {year} {2018})},\ \Eprint
  {http://arxiv.org/abs/1803.04961} {arXiv:1803.04961 [hep-ph]} \BibitemShut
  {NoStop}%
\bibitem [{\citenamefont {Czarnecki}\ \emph {et~al.}(2018)\citenamefont
  {Czarnecki}, \citenamefont {Marciano},\ and\ \citenamefont
  {Sirlin}}]{Czarnecki:2018okw}%
  \BibitemOpen
  \bibfield  {author} {\bibinfo {author} {\bibfnamefont {A.}~\bibnamefont
  {Czarnecki}}, \bibinfo {author} {\bibfnamefont {W.~J.}\ \bibnamefont
  {Marciano}}, \ and\ \bibinfo {author} {\bibfnamefont {A.}~\bibnamefont
  {Sirlin}},\ }\href@noop {} {\  (\bibinfo {year} {2018})},\ \Eprint
  {http://arxiv.org/abs/1802.01804} {arXiv:1802.01804 [hep-ph]} \BibitemShut
  {NoStop}%
\bibitem [{\citenamefont {Tang}\ \emph {et~al.}(2018)\citenamefont {Tang} \emph
  {et~al.}}]{Tang:2018eln}%
  \BibitemOpen
  \bibfield  {author} {\bibinfo {author} {\bibfnamefont {Z.}~\bibnamefont
  {Tang}} \emph {et~al.},\ }\href@noop {} {\  (\bibinfo {year} {2018})},\
  \Eprint {http://arxiv.org/abs/1802.01595} {arXiv:1802.01595 [nucl-ex]}
  \BibitemShut {NoStop}%
\bibitem [{\citenamefont {Pfützner}\ and\ \citenamefont
  {Riisager}(2018)}]{Pfutzner:2018ieu}%
  \BibitemOpen
  \bibfield  {author} {\bibinfo {author} {\bibfnamefont {M.}~\bibnamefont
  {Pfützner}}\ and\ \bibinfo {author} {\bibfnamefont {K.}~\bibnamefont
  {Riisager}},\ }\href@noop {} {\  (\bibinfo {year} {2018})},\ \Eprint
  {http://arxiv.org/abs/1803.01334} {arXiv:1803.01334 [nucl-ex]} \BibitemShut
  {NoStop}%
\bibitem [{\citenamefont {Plaster}\ \emph {et~al.}(2012)\citenamefont {Plaster}
  \emph {et~al.}}]{Plaster2012}%
  \BibitemOpen
  \bibfield  {author} {\bibinfo {author} {\bibfnamefont {B.}~\bibnamefont
  {Plaster}} \emph {et~al.} (\bibinfo {collaboration} {UCNA Collaboration}),\
  }\href {\doibase 10.1103/PhysRevC.86.055501} {\bibfield  {journal} {\bibinfo
  {journal} {Phys. Rev. C}\ }\textbf {\bibinfo {volume} {86}},\ \bibinfo
  {pages} {055501} (\bibinfo {year} {2012})}\BibitemShut {NoStop}%
\bibitem [{\citenamefont {Mendenhall}\ \emph {et~al.}(2013)\citenamefont
  {Mendenhall} \emph {et~al.}}]{Mendenhall2013}%
  \BibitemOpen
  \bibfield  {author} {\bibinfo {author} {\bibfnamefont {M.~P.}\ \bibnamefont
  {Mendenhall}} \emph {et~al.} (\bibinfo {collaboration} {UCNA
  Collaboration}),\ }\href {\doibase 10.1103/PhysRevC.87.032501} {\bibfield
  {journal} {\bibinfo  {journal} {Phys. Rev. C}\ }\textbf {\bibinfo {volume}
  {87}},\ \bibinfo {pages} {032501} (\bibinfo {year} {2013})}\BibitemShut
  {NoStop}%
\bibitem [{\citenamefont {Brown}\ \emph {et~al.}(2017)\citenamefont {Brown}
  \emph {et~al.}}]{Brown:2017mhw}%
  \BibitemOpen
  \bibfield  {author} {\bibinfo {author} {\bibfnamefont {M.~A.~P.}\
  \bibnamefont {Brown}} \emph {et~al.} (\bibinfo {collaboration} {UCNA}),\
  }\href@noop {} {\bibfield  {journal} {\bibinfo  {journal} {Phys. Rev. C (to
  be published)}\ } (\bibinfo {year} {2017})},\ \Eprint
  {http://arxiv.org/abs/1712.00884} {arXiv:1712.00884 [nucl-ex]} \BibitemShut
  {NoStop}%
\bibitem [{\citenamefont {Hickerson}\ \emph {et~al.}(2017)\citenamefont
  {Hickerson} \emph {et~al.}}]{Hickerson:2017fzz}%
  \BibitemOpen
  \bibfield  {author} {\bibinfo {author} {\bibfnamefont {K.~P.}\ \bibnamefont
  {Hickerson}} \emph {et~al.},\ }\href {\doibase 10.1103/PhysRevC.96.042501,
  10.1103/PhysRevC.96.059901} {\bibfield  {journal} {\bibinfo  {journal} {Phys.
  Rev.}\ }\textbf {\bibinfo {volume} {C96}},\ \bibinfo {pages} {042501}
  (\bibinfo {year} {2017})},\ \bibinfo {note} {[Addendum: Phys.
  Rev.C96,no.5,059901(2017)]},\ \Eprint {http://arxiv.org/abs/1707.00776}
  {arXiv:1707.00776 [nucl-ex]} \BibitemShut {NoStop}%
\bibitem [{\citenamefont {Holley}\ \emph {et~al.}(2012)\citenamefont {Holley}
  \emph {et~al.}}]{Holley2012}%
  \BibitemOpen
  \bibfield  {author} {\bibinfo {author} {\bibfnamefont {A.~T.}\ \bibnamefont
  {Holley}} \emph {et~al.},\ }\href {\doibase 10.1063/1.4732822} {\bibfield
  {journal} {\bibinfo  {journal} {Review of Scientific Instruments}\ }\textbf
  {\bibinfo {volume} {83}},\ \bibinfo {pages} {073505} (\bibinfo {year}
  {2012})},\ \Eprint {http://arxiv.org/abs/https://doi.org/10.1063/1.4732822}
  {https://doi.org/10.1063/1.4732822} \BibitemShut {NoStop}%
\bibitem [{\citenamefont {Saunders}\ \emph {et~al.}(2004)\citenamefont
  {Saunders} \emph {et~al.}}]{Saunders2003}%
  \BibitemOpen
  \bibfield  {author} {\bibinfo {author} {\bibfnamefont {A.}~\bibnamefont
  {Saunders}} \emph {et~al.},\ }\href {\doibase 10.1016/j.physletb.2004.04.048}
  {\bibfield  {journal} {\bibinfo  {journal} {Phys. Lett. B}\ }\textbf
  {\bibinfo {volume} {593}},\ \bibinfo {pages} {55} (\bibinfo {year}
  {2004})}\BibitemShut {NoStop}%
\bibitem [{\citenamefont {Morris}\ \emph {et~al.}(2002)\citenamefont {Morris}
  \emph {et~al.}}]{Morris2002}%
  \BibitemOpen
  \bibfield  {author} {\bibinfo {author} {\bibfnamefont {C.}~\bibnamefont
  {Morris}} \emph {et~al.},\ }\href {\doibase 10.1103/PhysRevLett.89.272501}
  {\bibfield  {journal} {\bibinfo  {journal} {Phys. Rev. Lett.}\ }\textbf
  {\bibinfo {volume} {89}},\ \bibinfo {pages} {272501} (\bibinfo {year}
  {2002})}\BibitemShut {NoStop}%
\bibitem [{\citenamefont {Saunders}\ \emph {et~al.}(2013)\citenamefont
  {Saunders} \emph {et~al.}}]{Saunders2013}%
  \BibitemOpen
  \bibfield  {author} {\bibinfo {author} {\bibfnamefont {A.}~\bibnamefont
  {Saunders}} \emph {et~al.},\ }\href {\doibase 10.1063/1.4770063} {\bibfield
  {journal} {\bibinfo  {journal} {Review of Scientific Instruments}\ }\textbf
  {\bibinfo {volume} {84}},\ \bibinfo {pages} {013304} (\bibinfo {year}
  {2013})},\ \Eprint {http://arxiv.org/abs/https://doi.org/10.1063/1.4770063}
  {https://doi.org/10.1063/1.4770063} \BibitemShut {NoStop}%
\bibitem [{\citenamefont {Plaster}\ \emph {et~al.}(2008)\citenamefont {Plaster}
  \emph {et~al.}}]{PlasterSCS2008}%
  \BibitemOpen
  \bibfield  {author} {\bibinfo {author} {\bibfnamefont {B.}~\bibnamefont
  {Plaster}} \emph {et~al.},\ }\href {\doibase 10.1016/j.nima.2008.07.143}
  {\bibfield  {journal} {\bibinfo  {journal} {Nucl. Instrum. Meth.}\ }\textbf
  {\bibinfo {volume} {A595}},\ \bibinfo {pages} {587} (\bibinfo {year}
  {2008})},\ \Eprint {http://arxiv.org/abs/0806.2097} {arXiv:0806.2097
  [nucl-ex]} \BibitemShut {NoStop}%
\bibitem [{\citenamefont {Ito}\ \emph {et~al.}(2007)\citenamefont {Ito} \emph
  {et~al.}}]{ItoMWPC2007}%
  \BibitemOpen
  \bibfield  {author} {\bibinfo {author} {\bibfnamefont {T.~M.}\ \bibnamefont
  {Ito}} \emph {et~al.},\ }\href {\doibase 10.1016/j.nima.2006.11.026}
  {\bibfield  {journal} {\bibinfo  {journal} {Nucl. Instrum. Meth.}\ }\textbf
  {\bibinfo {volume} {A571}},\ \bibinfo {pages} {676} (\bibinfo {year}
  {2007})},\ \Eprint {http://arxiv.org/abs/physics/0702085}
  {arXiv:physics/0702085 [PHYSICS]} \BibitemShut {NoStop}%
\bibitem [{\citenamefont {Morris}\ \emph {et~al.}(2009)\citenamefont {Morris}
  \emph {et~al.}}]{Morris2009}%
  \BibitemOpen
  \bibfield  {author} {\bibinfo {author} {\bibfnamefont {C.~L.}\ \bibnamefont
  {Morris}} \emph {et~al.},\ }\href {\doibase 10.1016/j.nima.2008.11.099}
  {\bibfield  {journal} {\bibinfo  {journal} {Nucl. Instrum.\ Meth.}\ }\textbf
  {\bibinfo {volume} {599}},\ \bibinfo {pages} {248} (\bibinfo {year}
  {2009})}\BibitemShut {NoStop}%
\bibitem [{\citenamefont {Rios}\ \emph {et~al.}(2011)\citenamefont {Rios} \emph
  {et~al.}}]{Rios2011105}%
  \BibitemOpen
  \bibfield  {author} {\bibinfo {author} {\bibfnamefont {R.}~\bibnamefont
  {Rios}} \emph {et~al.},\ }\href {\doibase
  https://doi.org/10.1016/j.nima.2010.12.098} {\bibfield  {journal} {\bibinfo
  {journal} {Nuclear Instruments and Methods in Physics Research Section A:
  Accelerators, Spectrometers, Detectors and Associated Equipment}\ }\textbf
  {\bibinfo {volume} {637}},\ \bibinfo {pages} {105 } (\bibinfo {year}
  {2011})}\BibitemShut {NoStop}%
\bibitem [{\citenamefont {Mendenhall}(2014)}]{Mendenhall2014}%
  \BibitemOpen
  \bibfield  {author} {\bibinfo {author} {\bibfnamefont {M.}~\bibnamefont
  {Mendenhall}},\ }\href@noop {} {Ph.D. thesis},\ \bibinfo  {school}
  {California Institute of Technology} (\bibinfo {year} {2014})\BibitemShut
  {NoStop}%
\bibitem [{\citenamefont {Agostinelli}\ \emph {et~al.}(2003)\citenamefont
  {Agostinelli} \emph {et~al.}}]{Geant4}%
  \BibitemOpen
  \bibfield  {author} {\bibinfo {author} {\bibfnamefont {S.}~\bibnamefont
  {Agostinelli}} \emph {et~al.} (\bibinfo {collaboration} {GEANT4}),\ }\href
  {\doibase 10.1016/S0168-9002(03)01368-8} {\bibfield  {journal} {\bibinfo
  {journal} {Nucl. Instrum. Meth.}\ }\textbf {\bibinfo {volume} {A506}},\
  \bibinfo {pages} {250} (\bibinfo {year} {2003})}\BibitemShut {NoStop}%
\bibitem [{\citenamefont {Lyons}(2008)}]{lyons2008}%
  \BibitemOpen
  \bibfield  {author} {\bibinfo {author} {\bibfnamefont {L.}~\bibnamefont
  {Lyons}},\ }\href {\doibase 10.1214/08-AOAS163} {\bibfield  {journal}
  {\bibinfo  {journal} {Ann. Appl. Stat.}\ }\textbf {\bibinfo {volume} {2}},\
  \bibinfo {pages} {887} (\bibinfo {year} {2008})}\BibitemShut {NoStop}%
\end{thebibliography}%
\end {document}